\documentstyle{article}
\begin{document}
\title{Graviton  noise:the Heisenberg picture}
\date{}

\author{ Z. Haba\\
Institute of Theoretical Physics, University of Wroclaw,\\
50-204 Wroclaw, Plac Maxa Borna 9,
Poland\\email:zbigniew.haba@uwr.edu.pl}\maketitle
\begin{abstract}We study   the geodesic deviation equation for a
quantum particle in a linearized quantum gravitational field.
Particle's Heisenberg equations of motion are treated as
stochastic  equations with a quantum noise. We explore the
stochastic equation beyond its local approximation as a
differential equation. We discuss the squeezed states resulting
from an inflationary evolution. We calculate the noise in the
thermal and squeezed states .
\end{abstract}

\section{Introduction}
For a long time an interaction of a massive particle with a
gravitational wave has been treated in analogy with an interaction
of a charged particle with an electromagnetic wave. The
electromagnetic wave can be represented as  a stream of photons.
The quantum description of light and its interaction with charged
particles has been well-developed and experimentally verified  in
quantum optics \cite{heitler}\cite{mandel}\cite{squizqed}. The
analogous quantum effects of gravity as proportional to the small
Newton constant $G$ would be much harder for detection
\cite{dyson}\cite{bing1}\cite{bing2}\cite{guer} . Quantum
gravitational processes  in an interaction of macroscopic masses
do sum up to the Newtonian force
\cite{feynmanbook}\cite{quantumforce} in a similar way as an
exchange of photons sums up to the Coulomb force between charged
particles. However, it seems to be difficult to detect the
corresponding quantum exchange phenomena (for some earlier work on
decoherence in particle-graviton interaction see
\cite{anasto}\cite{hu2}\cite{hu1}\cite{haba}\cite{hk}\cite{anastohu}\cite{wang}).
The detection of gravitational waves \cite{detectionwaves} has
been achieved owing to the immense gravitational energies emitted
during the merger of black holes. In order to detect quantum
gravitational phenomena from distant events  comparable energies
in such phenomena should be involved. It has been suggested by
Parikh, Wilczek and Zahariade
\cite{wilczek}\cite{wilczek1}\cite{wilczek2} (see also
\cite{soda1}\cite{soda2}) that an interaction with gravitons will
produce (besides the classical force) a quantum noise which can be
measured in gravitational wave detectors. In
refs.\cite{wilczek}\cite{wilczek1}\cite{wilczek2}
\cite{habaepj}\cite{chohu} the evolution of the particle density
matrix and the scattering probability in an environment of
gravitons have been calculated. In the semiclassical limit by
means of the Feynman-Vernon method \cite{feynman} a non-linear
stochastic differential equation with a noise from gravitons has
been derived (it is a stochastic perturbation of the equation of
geodesic deviation by the gravitational radiation reaction
\cite{gravitationbook}\cite{wald}).

In this paper we treat in detail the particle-graviton equations
of geodesic deviation as a quantum equation in the Heisenberg
picture. In quantum field theory such an approach is based on
Feldman-Yang equations \cite{schweber}. The Heisenberg equations
of motion can be solved perturbatively. Then, the correlation
functions can be calculated. For  particles in an environment of
an infinite number of gravitational modes in a Gaussian state
  the environment is producing an external Gaussian quantum noise \cite{gardiner}
  to
the particle system. In such a way a system of stochastic
equations is obtained. Quantum mechanics with dissipation
\cite{benguria1}\cite{benguria2} and linearized QED
\cite{ford1}\cite{ford2}have been studied in this way.  We show
that in the limit $\hbar\rightarrow 0$ and for small geodesic
deviations the stochastic differential equation  coincides with
the one derived by means of the Feynman-Vernon method in
refs.\cite{wilczek2} \cite{soda1}\cite{habaepj}. We investigate
the Heisenberg geodesic deviation equation beyond the above
mentioned approximations. We demonstrate that the radiation
reaction force does not depend on the state of the environment but
the noise does. We discuss the squeezed states resulting from an
inflationary evolution. We calculate the noise in thermal and
squeezed states. We show that the noise can be large in high
temperature and  in the squeezed states produced during inflation
\cite{grishchuk}. If we can detect primordial gravitational waves
then we may hope to detect the noise of the squeezed states. The
study of the geodesic deviation equation beyond the approximations
considered in \cite{wilczek2}\cite{soda1}\cite{habaepj} may be
relevant for the identification of the (quantum) gravitational
source.

The plan of the paper is the following. In secs.1-3 we review the
model of quantum geodesic deviation. In sec.4 we study the
geodesic deviation in a quantum thermal background calculating the
dependence of noise and backreaction on particle's space-time
location. In sec.5 we compute the quantum noise in a squeezed
state formed during inflation. In sec.6 we summarize the results.
In the Appendix we give some details of the calculations of the
non-local backreaction.
\section{Geodesic deviation}
The equation of geodesic deviation describes a relative motion of
two particles along the neighboring geodesics.  When $q^{\mu}$ is
the vector connecting the corresponding points of adjacent
geodesics $x^{\mu}$, then the acceleration of $q^{\mu}$ is
\cite{gravitationbook}
\begin{equation}
\frac{d^{2}q^{\mu}}{dt^{2}}=R^{\mu}_{\nu\sigma\rho}U^{\nu}U^{\sigma}q^{\rho},
\end{equation}where
$U^{\mu}=\frac{dx^{\mu}}{dt}$ and $R^{\mu}_{\nu\sigma\rho} $ is
the Riemanian curvature tensor. Eq.(1) can be considered as an
evolution equation for one particle of mass $m_{0}$ in the falling
frame of the other particle \cite{wilczek2}\cite{soda1}. We
consider eq.(1) on  the Minkowski background
$g_{\mu\nu}=\eta_{\mu\nu}+\lambda h_{\mu\nu}$ (where
$\eta_{\mu\nu}$ is the flat
 Minkowski metric,  $\lambda^{2}=8\pi
G$, $\mu,\nu=0,1,2,3$) in the non-relativistic approximation.
Then, $q=(t,{\bf q})$ and $U=(1,{\bf 0})$.  In a linear
approximation to the Riemannian tensor and in the
transverse-traceless gauge of $h$ eq.(1) reads
\cite{gravitationbook}($j,r=1,2,3$)
 \begin{equation}
\frac{d^{2}q^{j}}{dt^{2}}=\frac{1}{2}\lambda
\frac{d^{2}h^{jr}}{dt^{2}}q^{r}.
\end{equation}
In the  linearized gravity the Lagrangian describing the geodesic
deviation is quadratic in $h$
 (we set
the velocity of light $c=1$) \cite{wilczek2}\cite{soda1}
\begin{equation}
\begin{array}{l}
\int d{\bf x}{\cal L}=\frac{1}{8}\int d{\bf k} h^{*}_{\alpha}({\bf
k},t)((\frac{d}{dt})^{2}+k^{2})h_{\alpha}({\bf
k},t)+\frac{1}{2}m_{0}\frac{dq_{r}}{dt}\frac{dq_{r}}{dt}\cr-
\frac{1}{4}m_{0}\lambda (2\pi)^{-\frac{3}{2}}\int d{\bf
k}\exp(i{\bf kq})\frac{d^{2}h_{rl}}{dt^{2}}({\bf k})q_{r}q_{l}.
\end{array}\end{equation}
In eq.(3) $ h_{rl}$ is decomposed in the amplitudes
 $h_{\alpha}$ (where
 $\alpha=+,\times$ in the linear polarization) by means of the
 polarization tensors $e^{\alpha}_{rl}$  \cite{gravitation} \begin{equation}\begin{array}{l}
 h_{rl}({\bf
 x},t)=(2\pi)^{-\frac{3}{2}}\int d{\bf
k}(h_{\alpha}({\bf k})e^{\alpha}_{rl}\exp(-i{\bf
kx})+h_{\alpha}^{*}({\bf k})e^{\alpha}_{rl}\exp(i{\bf kx})).
\end{array}\end{equation}

 The gravitational Hamiltonian
$H=H_{+}+H_{\times}$ has the same form as for two independent
scalar fields $ h_{\alpha}$ \begin{equation} H=\frac{1}{2}\int
d{\bf k} \Big(-\frac{\delta}{\delta h^{\alpha}({\bf
k})}\frac{\delta}{\delta h^{\alpha}(-{\bf
k})}+k^{2}h^{\alpha}({\bf k})h^{\alpha}(-{\bf k})\Big).
\end{equation}

We look for Gaussian states of gravitons which are  solutions of
the Schr\"odinger equation  (with the Hamiltonian (5))
\begin{equation}
\psi_{t}^{g}(h)=A\exp\Big(\frac{i}{2\hbar}h\Gamma(t)h+\frac{i}{\hbar}J_{t}h\Big),
\end{equation}where $\Gamma$  is an operator defined by a bilinear form
$\Gamma(t,{\bf x}-{\bf y})$. Inserting $\psi_{t}^{g}$ in the
Schr\"odinger equation  with the Hamiltonian (5) we obtain
equations for $A,\Gamma,J$ (in Fourier space)
\begin{equation}\begin{array}{l}
i\hbar\partial_{t}\ln A=\frac{1}{2}\int d{\bf k} J({\bf k})J(-{\bf
k})-\frac{i\hbar}{2}\delta({\bf 0})\int d{\bf k}\Gamma({\bf k}),
\end{array}\end{equation}
where $\Gamma({\bf k})$ is the Fourier transform of $\Gamma({\bf
x})$,
\begin{equation}
\begin{array}{l}
\partial_{t}J=-\Gamma J,
\end{array}\end{equation}
\begin{equation}
\partial_{t}\Gamma=-\Gamma^{2}-k^{2} .
\end{equation}The term $\delta({\bf 0})$ in the normalization
factor(7)
 results from an infinite sum of vacuum oscillator energies. It could be made finite by
 a regularization of the Hamiltonian (5) but this is irrelevant for calculations of the expectation values
 (because the normalization factor cancels).
If we define
\begin{equation}
u(t)=\exp( \int^{t}ds \Gamma_{s}),
\end{equation}
then $\Gamma_{t}({\bf k})=u^{-1}\partial_{t}u$, where $u$
satisfies the equation
\begin{equation}
(\partial_{t}^{2}+k^{2})u(k)=0.
\end{equation}Eq.(11) has the general solution
\begin{equation}
u(t)=\sigma\cos(kt)+\delta\sin(kt).
\end{equation}
Another form of the solution is ( used by Guth and Pi \cite{guth})

\begin{equation}
u=\cos(kt+\alpha-i\gamma) , \end{equation}where the parameters
$\sigma,\delta,\alpha $ and $\gamma$ may depend on $k$.Then,
\begin{equation}\begin{array}{l}
\Gamma=-k\tan(kt+\alpha-i\gamma)= k\Big(i\sinh(2\gamma)-
\sin(kt+\alpha)\cos(kt+\alpha)\Big)\cr\times\Big(\cosh(2\gamma)
+\cos(2kt+2\alpha)\Big)^{-1}.
\end{array}\end{equation}
The covariance of the probability density
$\vert\psi_{t}^{g}\vert^{2}$ is equal to the inverse of

\begin{equation}
-i(\Gamma(t)-\Gamma^{*}(t))=2k\sinh(2\gamma)\Big(\cosh(2\gamma)
+\cos(2kt+2\alpha)\Big)^{-1}.
\end{equation}
The fluctuations of $h$ are large if $\gamma$ is  small. It
follows from eq.(13) that during the evolution described by the
Hamiltonian (5) only the phase is changing $\alpha\rightarrow
\alpha +kt$. We can express $\alpha$ and $\delta$ in eq.(12) by
$\alpha$ and $\gamma$ of eq.(13). The form (13) of the solution of
eq.(11) is useful in order to express the squeezing by a small
value of $\gamma$.

\section{Heisenberg equations of motion}
 The equation for the gravitational field resulting from the
Lagrangian (3) is
\begin{equation}
\frac{d^{2}h^{rl}({\bf k})}{dt^{2}}+k^{2}h^{rl}({\bf
k})=(2\pi)^{-\frac{3}{2}}\lambda f^{rl}\exp(-i{\bf kq}(t)),
\end{equation}where
\begin{equation}
f^{rl}=\frac{m_{0}}{2}\frac{d^{2}}{dt^{2}}q^{r}q^{l}.
\end{equation}
 The  solution of eq.(16) in the transverse-traceless
gauge for $t\geq t_{0}$ (we assume that when $t\leq t_{0}$ then
 $h_{rl}$ is a free wave) is
\begin{equation}\begin{array}{l} h_{rl}({\bf
k})=h_{rl}^{w}+h_{rl}^{f}+h_{rl}^{I}\equiv
h_{rl}^{w}+e_{rl}^{\alpha}(h_{0}^{\alpha}\cos(kt)+k^{-1}\Pi_{0}^{\alpha}\sin(kt))+
\cr\lambda (2\pi)^{-\frac{3}{2}}
\Lambda_{rl;mn}\int_{t_{0}}^{t}k^{-1}\sin(k(t-t^{\prime}))f_{mn}(t^{\prime})\exp(-i{\bf
kq}(t^{\prime}))dt^{\prime},
\end{array}\end{equation}
here $\Lambda$ projects to the transverse-traceless gauge
\begin{equation}\begin{array}{l}
2\Lambda_{ij;mn}(\frac{{\bf
k}}{k})=2e^{\alpha}_{ij}e^{\alpha}_{mn}=(\delta_{im}-k^{-2}k_{i}k_{m})
(\delta_{jn}-k^{-2}k_{j}k_{n})\cr+(\delta_{in}-k^{-2}k_{i}k_{n})
(\delta_{jm}-k^{-2}k_{j}k_{m})-\frac{2}{3}(\delta_{ij}-k^{-2}k_{i}k_{j})
(\delta_{nm}-k^{-2}k_{n}k_{m}).\end{array}
\end{equation}
$h_{rl}^{w}$ describes the classical wave (possibly one mode of
it) and
\begin{displaymath}
h^{f}_{rl}(t,{\bf x}) =(2\pi)^{-\frac{3}{2}}\int d{\bf
k}\exp(i{\bf kx})h^{f}_{rl}({\bf k},t)
\end{displaymath}
is a superposition of quantum free modes decomposed in eq.(18)
into $h_{0}^{\alpha}$ and $\Pi_{0}^{\alpha}$ as quantum initial
conditions. $h_{rl}^{I}$ is the gravitational field created by the
particle motion. $h_{rl}^{f}$ (as well as $h_{rl}^{w}$) satisfy
the homogeneous equation
\begin{displaymath}
\frac{d^{2}h_{rl}^{f}({\bf k})}{dt^{2}}+k^{2}h_{rl}^{f}({\bf
k})=0.
\end{displaymath} The equation of motion for the coordinate ${\bf q}$ is

\begin{equation}\begin{array}{l}
\frac{d^{2}q_{r}}{dt^{2}}=\frac{\lambda}{2}(2\pi)^{-\frac{3}{2}}\int
d{\bf k}\exp(i{\bf kq}(t))(\frac{d^{2}h^{w}_{rl}({\bf
        k})}{dt^{2}}+\frac{d^{2}h^{f}_{rl}({\bf
        k})}{dt^{2}})q^{l}(t) \cr
+\frac{\lambda}{2}(2\pi)^{-\frac{3}{2}}\int d{\bf k}\exp(i{\bf
kq}(t))\frac{d^{2}h_{rl}^{I}({\bf k})}{dt^{2}}q^{l}(t).
\end{array}\end{equation}
We insert the gravitational field from eq.(18) into eq.(20). Then,
\begin{equation}
\frac{d^{2}q_{r}}{dt^{2}}=\frac{\lambda}{2}\frac{d^{2}h^{w}_{rl}({\bf
        q})}{dt^{2}}q^{l}(t) +F_{r}({\bf q},t)+N_{rl}(t,{\bf
q})q^{l},
\end{equation}
where F is a non-linear interaction (the backreaction  discussed
in the next section) resulting from the interaction of the mass
$m_{0}$ with its own gravitational field. The noise $N_{rl}(t,{\bf
q})$ is expressed by the initial values of the canonical (free)
variables $h^{\alpha}({\bf k}) $ and $\Pi^{\alpha}({\bf k})$
\begin{equation}\begin{array}{l} N^{rl}(t,{\bf q})=\int d{\bf k}
N^{rl}({\bf k},{\bf
q},t)\cr=-\frac{\lambda}{2}(2\pi)^{-\frac{3}{2}}\int d{\bf
k}\exp(i{\bf kq}(t)) k^{2}e^{\alpha}_{rl}
(h_{0}^{\alpha}\cos(kt)+k^{-1}\Pi_{0}^{\alpha}\sin(kt)).
\end{array}\end{equation}
\section{Stochastic equations in the thermal environment}
Let us first consider ${\bf q}\simeq 0$ in eq.(22)(then we do not
need to care about the non-commutativity of ${\bf q}(t)$). We
denote $N_{rl}(t)\equiv N_{rl}(t,{\bf 0})$. Then

\begin{equation}\begin{array}{l}
N^{rl}(t)=\int d{\bf k} N^{rl}(k,{\bf
0},t)=-\frac{\lambda}{2}(2\pi)^{-\frac{3}{2}}\int d{\bf k}
k^{2}e^{\alpha}_{rl}
(h_{0}^{\alpha}\cos(kt)+k^{-1}\Pi_{0}^{\alpha}\sin(kt)).
\end{array}\end{equation} Let us  assume that $h^{\alpha}({\bf k}) $ and $\Pi^{\alpha}({\bf
k})$ are distributed according to the classical Gibbs distribution
\begin{equation}
d\Pi_{0}^{\alpha}dh_{0}^{\alpha}\exp\Big(-\frac{\beta}{2}\int
d{\bf k}(\vert\Pi_{\alpha}\vert^{2}+k^{2}\vert
h_{\alpha}\vert^{2})\Big).
\end{equation} here $\beta=\frac{1}{k_{B}T}$ where $T$ is  the temperature and $k_{B}$ is the
Boltzmann constant. Then,
\begin{equation}
<\overline{h}_{0}^{\alpha}({\bf k})h_{0}^{\alpha}({\bf
k}^{\prime})>=\beta^{-1}k^{-2}\delta({\bf k}-{\bf k}^{\prime}),
\end{equation}and\begin{equation}
<\overline{\Pi}_{0}^{\alpha}({\bf k})\Pi_{0}^{\alpha}({\bf
k}^{\prime})=\beta^{-1}\delta({\bf k}-{\bf k}^{\prime}).
\end{equation}The noise (23) has the correlations calculated in
\cite{habaepj} \begin{equation}\Big< N^{rl}(t)
N^{mn}(t^{\prime})\Big>=<\Lambda_{rl:mn}>\frac{\lambda^{2}}{4\pi}
\beta^{-1}\partial_{t}^{2}\partial_{t^{\prime}}^{2}\delta(t-t^{\prime}),
\end{equation}where
the angular average of $\Lambda$ over ${\bf k}k^{-1}$ is
\begin{equation}
\frac{1}{4\pi}<\Lambda_{ij;mn}>=\frac{1}{5}(\delta_{im}\delta_{jn}+\delta_{in}\delta_{jm})
-\frac{2}{15}\delta_{ij}\delta_{nm}.
\end{equation}
When ${\bf q}$ is taken into account (but non-commutativity of
${\bf q}$'s is ignored)

\begin{equation}\begin{array}{l}
\Big< \overline{N}^{rl}({\bf q},t) N^{mn}({\bf
q},t^{\prime})\Big>\cr=\frac{\lambda^{2}}{4}\beta^{-1}(2\pi)^{-3}\int
d{\bf k}k^{2}\Lambda_{rl:mn}\cos(k(t-t^{\prime}))\exp(i{\bf
kq}(t^{\prime})-i{\bf kq}(t)) .
\end{array}\end{equation}
In eq.(29) we can calculate the angular average over $k^{-1}{\bf
k}$ using the formula \begin{equation}\begin{array}{l} \int
d\Omega\Lambda^{rl;mn}({\bf k}k^{-1}) \exp(i{\bf k}{\bf
x})=4\pi\Lambda^{rl;mn}(k^{-1}\nabla_{\bf x}) (k\vert {\bf
x}\vert)^{-1}\sin(k\vert {\bf x}\vert)
\end{array}\end{equation} where $d\Omega$ is an average over the
spherical angle .

We can express the integral (29) in configuration space performing
the $k$-integral of trigonometric functions using the result (28)
for the average   $<\Lambda_{rl:mn}>$ (19) over the spherical
angle and differentiating the formula
\begin{equation}
\int_{0}^{\infty}dk k^{-1}\sin(ku)=\frac{\pi}{2}\epsilon(u)
\end{equation} over $u$
where $\epsilon(u)$ is an antisymmetric function such that
$\epsilon(u)=1$ for $u>0$. The result of calculations gives a long
formula but the main outcome is that   noise is concentrated on
the light cone.

 The quantum Bose-Einstein distribution is (for general discussion
 of quantum noise see \cite{gardiner}\cite{welton}\cite{noise})
\begin{equation}
<(h^{\alpha}({\bf k}))^{+}h^{\alpha}({\bf
k}^{\prime})>=\frac{1}{2}\hbar k^{-1}\coth(\frac{1}{2}\hbar\beta
k))\delta({\bf k}-{\bf k}^{\prime})
\end{equation}and\begin{equation}
<(\Pi^{\alpha}({\bf k}))^{+}\Pi^{\alpha}({\bf
k}^{\prime})>=\frac{1}{2}\hbar k\coth(\frac{1}{2}\beta
k))\delta({\bf k}-{\bf k}^{\prime}).
\end{equation}
Then, for the noise we obtain
\begin{equation}\begin{array}{l}
\frac{1}{2}\Big< N^{mn}({\bf q},t^{\prime})N^{rl}({\bf
q},t)+N^{rl}({\bf q},t)N^{mn}({\bf
q},t^{\prime})\Big>\cr=\frac{\lambda^{2}}{4}(2\pi)^{-3}\int d{\bf
k})k^{4}\Lambda_{rl:mn}\frac{1}{2}\hbar
k^{-1}\coth(\frac{1}{2}\hbar\beta
k))\cos(k(t-t^{\prime}))\exp(i{\bf kq}(t^{\prime})-i{\bf kq}(t)) .
\end{array}\end{equation}
In the quantum case the operators $N^{rl}(t,{\bf q})$ do not
commute
\begin{equation}\begin{array}{l}
[N^{rl}(t,{\bf q}),N^{mn}(t^{\prime},{\bf
q}^{\prime})]\cr=i\hbar\frac{\lambda^{2}}{4}(2\pi)^{-3}\int d{\bf
k}k^{3}\Lambda^{rl:mn}\sin(k(t-t^{\prime}))\exp(i{\bf kq}(t)-i{\bf
kq}(t^{\prime})).\end{array}
\end{equation}
From the uncertainty relation we can conclude that the noise
cannot be arbitrarily small what is bringing difficulties in the
precision of measurements \cite{welton}\cite{noise} but according
to \cite{wilczek}\cite{wilczek2} this is just the large noise
which is interesting for investigation in gravitation wave
experiments.

The non-linear force resulting from the graviton environment is
\begin{equation}\begin{array}{l}
F^{r}=\frac{1}{2}\lambda^{2}(2\pi)^{-3}\int d{\bf k}\exp(i{\bf
kq}(t)) q^{l}(t)
\Lambda_{rl;mn}\cr\partial_{t}^{2}\int_{t_{0}}^{t}k^{-1}\sin(k(t-t^{\prime}))\exp(-i{\bf
kq}(t^{\prime}))f_{mn}(t^{\prime})dt^{\prime}\cr
=\frac{1}{2}\lambda^{2}(2\pi)^{-3}\int d{\bf k}\exp(i{\bf
kq}(t))q^{l}(t) \cr
\Lambda_{rl;mn}(\int_{t_{0}}^{t}\partial_{t}\cos(k(t-t^{\prime}))\exp(-i{\bf
kq}(t^{\prime})f_{mn}(t^{\prime})dt^{\prime} +f_{mn}(t)\exp(-i{\bf
kq}(t)).
\end{array}\end{equation}
We write the last factor as\begin{equation}\begin{array}{l}
(\int_{t_{0}}^{t}\partial_{t}\cos(k(t-t^{\prime}))\exp(-i{\bf
kq}(t^{\prime}))f_{mn}(t^{\prime})dt^{\prime}
+f_{mn}(t)\exp(-i{\bf
kq}(t))\cr=(-\int_{t_{0}}^{t}\partial_{t^{\prime}}\cos(k(t-t^{\prime}))\exp(-i{\bf
kq}(t^{\prime}))f_{mn}(t^{\prime})dt^{\prime}
+f^{mn}(t)\exp(-i{\bf
kq}(t))\cr=\int_{t_{0}}^{t}\cos(k(t-t^{\prime}))\Big(dt^{\prime}\partial_{t^{\prime}}\Big(f_{mn}(t^{\prime})\exp(-i{\bf
kq}(t^{\prime}))\Big)\cr+\cos(k(t-t_{0}))f_{mn}(t_{0})\exp(-i{\bf
kq}(t_{0}))\cr=\int_{t_{0}}^{t}\cos(k(t-t^{\prime}))\partial_{t^{\prime}}\Big(f_{mn}(t^{\prime})\exp(-i{\bf
kq}(t^{\prime}))dt^{\prime}\Big)\cr=\int_{t_{0}}^{t}dt^{\prime}\cos(k(t-t^{\prime}))\exp(-i{\bf
kq}(t^{\prime}))(\partial_{t^{\prime}}f_{mn}(t^{\prime})-i{\bf
k}\frac{d{\bf q}}{dt^{\prime}}),
\end{array}\end{equation}
where  we assumed $f_{mn}(t_{0})=0$. In the approximation
$\exp(-i{\bf kq}(t^{\prime}))\simeq 1$ the integral (37) is
\begin{equation}\begin{array}{l}
\int_{t_{0}}^{t}\int dk
k^{2}\cos(k(t-t^{\prime}))\partial_{t^{\prime}}f_{mn}(t^{\prime})dt^{\prime}\cr=-2\pi\int_{t_{0}}^{t}\partial_{t^{\prime}}^{2}
\delta(t-t^{\prime})\partial_{t^{\prime}}f_{mn}(t^{\prime})dt^{\prime}
=2\pi\partial_{t}^{3}f_{mn}(t).\end{array}
\end{equation}
Then, we can write eq.(21) in the form ( we omit the classical
wave $h^{w}$)
\begin{equation}\begin{array}{l}
\frac{d^{2}q^{r}}{dt^{2}}=-\frac{\lambda^{2}}{5\pi}(\delta_{rm}\delta_{ln}-\frac{1}{3}\delta_{rl}\delta_{mn})
q^{l}\partial_{t}^{3}f^{mn} +N^{rl}(t)q_{l}.
\end{array}\end{equation}
This is the perturbation by noise obtained in
\cite{wilczek}\cite{wilczek2}\cite{habaepj} of the gravitational
backreaction equation of ref.\cite{wald}.

If in eq.(36) we take $\exp(-i{\bf kq}(t^{\prime}))$ into account
then we obtain an integro-differential equation ( we cannot get
rid of the $t^{\prime}$ integral in eq.(36) but the ${\bf k}$
integral can be calculated, see the Appendix). The force is
non-local in time, but its exact form may be relevant for an
interpretation of particle motion. The gravitational wave
detectors have a finite resolution range in frequency as well as
in space and time. For this reason it is useful to have
expressions for the noise correlations  both in the frequency
domain and in the space-time .

 In the
calculations (35) and (36) we have ignored the non-commutativity
of ${\bf q}(t)$ and ${\bf q}(t^{\prime})$. We have
\begin{equation}\begin{array}{l} \exp(i{\bf kq}(t))\exp(-i{\bf kq}(t^{\prime}))=
\exp(i{\bf kq}(t))-i{\bf kq}(t^{\prime}))
\cr\times\exp\Big(\frac{1}{2}[{\bf kq}(t)),{\bf
kq}(t^{\prime})]\Big)\Big(1+O(\hbar^{2})\Big)
\end{array}\end{equation} Eq.(39) holds true in the limit
$\hbar\rightarrow 0$ and with the approximation that ${\bf
q}\simeq{\bf 0}$. In \cite{wilczek2} the "particle" is
macroscopic. Then ${\bf q}$ is classical. However, eq.(21) is
applicable to quantum particles as well. We could have gravitons
interacting with molecules or crystals. The variation of
molecule's length or crystal's size could be transmitted to an
interferometer.
 We can
allow large $q$ as in eqs.(29) and (36) but such equations are
reliable only till the order $\hbar$, because at higher orders of
$\hbar$ we would need to solve the Heisenberg equations of motion
(21) (it seems possible only in a perturbative expansion in
$\hbar$)till the higher order in $\hbar$ and calculate the
commutators ${\bf q}(t)$ and ${\bf q}(t^{\prime})$ (as in
eq.(40)).

\section{The noise from the squeezed states of (inflationary)
gravitons} In this section we study tensor perturbations
(gravitational waves) evolving  during the inflationary epoch and
subsequently reaching us in a flat  Minkowski space. We consider
the flat expanding metric
\begin{equation}
ds^{2}=g_{\mu\nu}dx^{\mu}dx^{\nu} =dt^{2}-a^{2}d{\bf x}^{2}.
\end{equation}
It is useful to introduce the conformal time $\tau$
\begin{displaymath}
\tau=\int^{t}a(s)^{-1}ds.
\end{displaymath}
We decompose $h_{rl}$ into polarization components $h^{\alpha}$
 (as in eq.(4) for $a=1$)  $h_{rl}=a^{-1}e^{\alpha}_{rl}h^{\alpha}$. Then the Hamiltonian is
 \cite{mukhanovbook}

\begin{equation}H=\frac{1}{2}\int d{\bf
x}\Big((\Pi^{\alpha})^{2}-h^{\alpha}(\triangle+a^{\prime\prime}a^{-1}
 )h^{\alpha}\Big),
\end{equation}
where $\Pi^{\alpha}=-i\hbar\frac{\delta}{\delta h^{\alpha}}$ is
the canonical momentum. The solution of the Schr\"odinger equation
\begin{displaymath}i\hbar\partial_{\tau}\Psi=H\Psi
\end{displaymath}
has  the Gussian form (6)
 if $\Gamma$
satisfies the equation\begin{equation}
\partial_{\tau}\Gamma+\Gamma^{2}+(k^{2}-a^{\prime\prime}a^{-1}
 )\Gamma=0.\end{equation}
 As in eq.(10) we express $\Gamma$ in the form
 $\Gamma=u^{-1}\partial_{\tau}u$.
Then, $u$ is a solution of the  equation
\begin{equation}
\partial_{\tau}^{2}u +(k^{2}-a^{\prime\prime}a^{-1}
 )u=0.\end{equation}
 In the inflationary era described by an exponential expansion  $a=\exp(H_{0}t)$ (the cosmic time
 $t$ is related to $\tau$ in eqs.(43)-(44)    as
$\tau=-H_{0}^{-1}\exp(-H_{0}t)$ , where $H_{0}$ is the Hubble
constant) eq.(44) reads

\begin{equation}
\partial_{\tau}^{2}u +(k^{2}-2\tau^{-2} )u=0.\end{equation}
The general solution of eq.(45) (analogous to the solution (12) in
the Minkowski space) is
\begin{equation} u=\sigma\cos(k\tau)+\delta\sin(k\tau)
-\sigma(k\tau)^{-1}\sin(k\tau)+\delta (k\tau)^{-1}\cos(k\tau).
\end{equation}
 The solution which behaves as a plane wave $\exp(ik\tau)$ for
$\tau\rightarrow \infty$ results from the choice $\sigma=1$ and
$\delta=i$. Then, at the end $\tau_{e}$ of inflation ( for a small
$k\tau_{e}$)
\begin{equation}
\Gamma_{e}\simeq
(ik^{3}\tau_{e}^{2}-\tau_{e}^{-1})(1+k^{2}\tau_{e}^{2})^{-1}.
\end{equation}
The squeezing and decoherence during inflation has been discussed
earlier in \cite{grishchuk}\cite{aa}\cite{star1}\cite{star2}(the
$k^{-3}$ covariance is characteristic of the exponential expansion
\cite{bunch}\cite{abbott}). We take
$\psi_{0}^{e}=A\exp(\frac{i}{2\hbar}h^{\nu}\Gamma_{e}h^{\nu})$ as
the initial wave function for our graviton detection experiment.
From its derivation it is clear that the non-linear term for the
particle geodesic deviation equation (21) does not depend on the
graviton wave function (although this term may have the non-local
correction in comparison to eq.(39) as discussed in the Appendix).
In eq.(21) the noise term (its correlation function) depends on
the quantum state of the graviton.  We have discussed the
evolution of Gaussian wave packets in various epochs in
\cite{habauniverse}. We concluded that the squeezing described by
small $i(\Gamma_{e}-\Gamma_{e}^{*})$ remains small ($\simeq
k^{3}$) during the radiation and baryonic eras. According to
eq.(14) during the Minkowski space evolution only the phase
$\alpha$ is changing.

 We calculate now the noise
correlation functions in the Gaussian state $\psi_{0}^{e}$. For
this purpose we may use the expression for the noise (22) in terms
of the initial field and the canonical momentum ( the noise
correlation functions are calculated in a different
parametrization of squeezed states in
\cite{wilczek2}\cite{soda1}\cite{hertzberg}). Then (if $J=0$) the
noise has the correlations
\begin{equation}\begin{array}{l}
(\psi_{0}^{e},N^{rl}(t,{\bf q})N^{mn}(t^{\prime},{\bf
q}^{\prime})\psi_{0}^{e})=\frac{\lambda^{2}}{4}(2\pi)^{-3}\int
d{\bf k}k^{4}\exp(i{\bf k}({\bf q}(t)-{\bf
q}(t^{\prime}))\Lambda^{rl;mn} \cr\Big(i\hbar
(\Gamma_{e}-\Gamma_{e}^{*})^{-1}(\cos(kt)+k^{-1}\Gamma_{e}\sin(kt))
(\cos(kt^{\prime})+k^{-1}\Gamma_{e}\sin(kt^{\prime})) \cr-i\hbar
k^{-1}\Big(\sin(kt)\cos(kt^{\prime})+\Gamma_{e}k^{-1}\sin(kt)\sin(kt^{\prime})\Big)\Big)

\end{array}\end{equation}
 If we make the
usual  assumption that $k\tau_{e}$ is small and  use the
approximation $\exp(i{\bf kq})\simeq 1$ then
$\Gamma_{e}-\Gamma_{e}^{*}\simeq 2ik^{3}\tau_{e}^{2}$. Hence,
\begin{equation}\begin{array}{l}
(\psi_{0}^{e},N^{rl}(t,{\bf 0})N^{mn}(t^{\prime},{\bf
0})\psi_{0}^{e})  \simeq\cr
\tau_{e}^{-2}\frac{\lambda^{2}}{4}(2\pi)^{-3}2\pi<\Lambda^{rl;mn}>\int
dkk^{3} \hbar k^{-3} \cos(kt)\cos(kt^{\prime})\cr=
\tau_{e}^{-2}\frac{\lambda^{2}}{4}(2\pi)^{-3}2\pi^{2}<\Lambda^{rl;mn}>
\partial_{t}\partial_{t^{\prime}}\cr\times\Big((\partial_{t}-\partial_{t^{\prime}})\delta(t-t^{\prime})
+(\partial_{t}+\partial_{t^{\prime}})\delta(t+t^{\prime})\Big)\end{array}\end{equation}
which can be large for a small value of $\tau_{e}$.

 The aim of the stochastic
geodesic deviation equation in
refs.\cite{wilczek}\cite{wilczek2}\cite{soda1} is to describe the
motion of the arms of the interferometer. In such a case the
classical (more precisely a semi-classical) limit (39) of eq.(21)
may be sufficient. However, eq.(21) can describe quantum objects
(large molecules, crystals) as well. The interferometers can be
sensitive to their change of size. In such a case, besides the
noise, the quantum evolution of the coordinate ${\bf q}$ may be
relevant. It is not simple to solve the non-linear quantum
Heisenberg equation (21). One can do it in perturbation expansion
in $\lambda$ and $\hbar$. For the purpose of illustration let us
consider an oscillator of frequency $\omega$, i.e.,
 we add to eq.(39) the force $-m_{0}\omega^{2}q^{r}$. Then,
 eq.(39) reads
\begin{equation}\begin{array}{l}
\frac{d^{2}q^{r}}{dt^{2}}+\omega^{2}q^{r}=-\frac{\lambda^{2}}{5\pi}(\delta_{rm}\delta_{ln}-\frac{1}{3}\delta_{rl}\delta_{mn})
q^{l}\partial_{t}^{3}f^{mn} +\lambda \tilde{N}^{rl}(t)q_{l}.
\end{array}\end{equation}
Here we write $N^{rl}(t)=\lambda \tilde{N}^{rl}(t) $ where
$\tilde{N}^{rl}(t)$ does not depend on $\lambda$. We look for
perturbative solutions of eq.(50) in an expansion in
$\lambda=\sqrt{8\pi G}$. Let $q_{0}^{r}$ be the solution of the
harmonic oscillator (zeroth order in $\lambda$). At the first
order in $\lambda$
\begin{equation} \frac{d^{2}q_{1}^{r}}{dt^{2}}+\omega^{2}q_{1}^{r}=
\tilde{N}^{rl}(t)q_{0}^{l}. \end{equation} At order $\lambda^{2}$
we obtain
\begin{equation}\begin{array}{l}
\frac{d^{2}q_{2}^{r}}{dt^{2}}+\omega^{2}q_{2}^{r}=-\frac{64}{15\pi}
\omega^{4}m_{0}q_{0}^{r}q_{0}^{n}\frac{dq_{0}^{n}}{dt}
+\tilde{N}^{rl}(t)q_{1}^{l}.
\end{array}\end{equation}
Eq.(52) shows the friction term $\gamma^{rn}\frac{dq^{n}}{dt}$ in
Newton equation. The friction matrix $\gamma$ has the non-zero
eigenvalue $\frac{512 Gm_{0}}{15}{\bf q}^{2}\omega^{4}$.
 The ratio of the gravitational friction
to the electromagnetic one is approximately  $ {\bf
q}^{2}\omega^{2}m_{0} G e^{-2}$ where $e$ is the electric charge
and $G$ is the Newton constant. For the electron $m_{0} G
e^{-2}\simeq 10^{-42}$. Hence, the gravitational friction  is
unmeasurable unless $\omega^{2} {\bf q}^{2} $ is large. As pointed
out in \cite{wilczek}\cite{wilczek2} this is the noise in eq.(39)
which can be measurable if $\tau_{e}$ is small. The friction is
related to the width of the spectral line \cite{heitler} so the
gravitational effect on spectral lines is not observable  as
discussed from another point of view in \cite{bing1}\cite{bing2}.

\section{Summary and conclusions}
We have studied in some detail the quantum geodesic deviation
equation in the Heisenberg picture beyond the earlier
semi-classical approximations. We reveal the dependence of the
backreaction force and of the quantum noise on the particle
position $q$ when $kq$ is not negligible where $k$ describes the
graviton's wave number distribution in a quantum graviton state.
For primordial gravitons coming from the inflationary era the main
contribution to the graviton's probability distribution comes from
small $k$ (then $kq$ is small). However, for thermal gravitons
large $k$ may be important (depending on the relevant values of
$q$). The calculations of the quantum noise correlation functions
considered in this paper may be useful for a determination of the
source of the gravitational waves if they have a quantum origin .
\section{Appendix: the non-local backreaction}

 The force in eq.(36) is
\begin{equation}\begin{array}{l}
F^{r}=\frac{1}{2}\lambda^{2}(2\pi)^{-3}q^{l}(t)\int d{\bf
k}\exp(i{\bf kq}(t)-i{\bf kq}(t^{\prime})) \cr
\Lambda_{rl;mn}(\int_{t_{0}}^{t}dt^{\prime}\cos(k(t-t^{\prime}))(\partial_{t^{\prime}}f_{mn}(t^{\prime})-i{\bf
k}\frac{d{\bf q}}{dt^{\prime}}).
\end{array}\end{equation}
We can perform the angular integral using eq.(30). Then
\begin{equation}\begin{array}{l}
F_{I}^{r}=\frac{1}{2}\lambda^{2}(2\pi)^{-3}q^{l}(t)4\pi\int_{0}^{\infty}dk
k ^{2}\int_{t_{0}}^{t}dt^{\prime}\cos(k(t-t^{\prime}))
\partial_{t^{\prime}}f_{mn}(t^{\prime})\Lambda_{rl;mn}(\frac{\nabla_{\bf q}}{k})
\cr\times\Big(k^{-1}\sin(k\vert {\bf q}(t)-{\bf
q}(t^{\prime})\vert)\vert {\bf q}(t)-{\bf q}(t^{\prime})\vert^{-1}
\cr+\cos(k\vert {\bf q}(t)-{\bf q}(t^{\prime})\vert)\vert {\bf
q}(t)-{\bf q}(t^{\prime})\vert^{-2}{\bf q}(t^{\prime})({\bf
q}(t)-{\bf q}(t^{\prime}))\cr-k^{-1}\sin(k\vert {\bf q}(t)-{\bf
q}(t^{\prime})\vert)\vert {\bf q}(t)-{\bf
q}(t^{\prime})\vert^{-3}{\bf q}(t^{\prime})({\bf q}(t)-{\bf
q}(t^{\prime}))\Big)
\end{array}\end{equation}
Performing the differentiation $\Lambda_{rl;mn}(\frac{\nabla_{\bf
q}}{k})$ we obtain a trigonometric function multiplied by an
integer power of $k$. The $k$-integral can be calculated using the
representation (31) of the $\epsilon $ function (and derivatives
of it). The integral representation may be useful for a study of
non-local effects in the solutions of the quantum geodesic
deviation equation.

 \end{document}